Phillip H. Roth

# *Formalizing Informal Communication: An Archaeology of the Early Pre-Web Preprint Infrastructure at CERN*

**Abstract:**

This article deals with the early development of preprint communication in high-energy physics (HEP), specifically with how preprint communication was formalized in the early 1960s at the European Organization for Nuclear Research (CERN). It employs a sociological conception of infrastructures to ask which practices and technologies of communication structured the use of preprints in HEP at the time and subsequently solidified into the research community's communication and information system. The text conducts an archaeology of the early preprint infrastructure along the lines of three systematic-historical explorations: 1) the use of preprints as media to privately and informally share practical instructions and theoretical tools in the fast-moving current of postwar theoretical physics; 2) the institutional and organizational context of library and documentation work at CERN around mid-century, which acted as a backdrop for creating the preprint infrastructure; and 3) the actual formalization of preprint communication into an information system at the CERN library in the early 1960s, which treated preprints as public "current awareness tools" for the benefit of the whole HEP community, informing members of recent progress in the field. The article concludes with a discussion of what we can learn from the pre-Web history of the preprint communication in HEP about current understandings of scientific communication.

## 1. Introduction

Preprints have received quite some attention as a subject of analysis in recent years, especially in relation to online repositories. In common understanding, preprints are preliminary versions of scientific papers made available to peers, sometimes in the form of articles accepted for publication in an academic journal and sometimes in the form of texts not submitted for publication. Networked communication technologies and the emergence of the World Wide Web in the early 1990s have allowed researchers to share findings with colleagues through practices of so-called “self-archiving,” where papers are made freely available on platforms for others to download and read (Bohlin 2004, Borrelli 2022, Gunnarsdóttir 2005, Taubert 2021).[1] Such practices gained widespread attention within academia when during the Covid-19 pandemic biomedical scientists, epidemiologists, and clinicians shared their discoveries through preprints at an “unprecedented pace” (Watson 2022). Close to one hundred different preprint servers exist today – some institution-led, some owned by commercial publishers, some covering special disciplines, and some open to any field for self-archiving.[2]

In social studies of science, scholars have analyzed how preprint communication infrastructures like that of the online repository arXiv.org have complemented “traditional publication practices” (Gunnarsdóttir 2005). In physics, astronomy, and mathematics, where the repository has been employed by researchers since the 1990s, the routines of scientific authors and readers are organized around the arXiv infrastructure and researchers are included in the disciplinary communication system through practices of self-archiving. Providing important epistemic and social resources to pursue their day-to-day work, in mathematics and astronomy, for instance, “repositories register claims for new results and findings” (Taubert 2021: 190), while in high-energy physics, “preprint archives have assumed a central role within the field as the space where belonging and credit are constructed and confirmed.” (Delfanti 2016: 642) Moreover, as Reyes-Galindo (2016: 600) explains, the “sophisticated combination of software filtering, moderator scrutiny and peer-exposure in arXiv” has led to the emergence of gate-keeping functions comparable to that of journal editorial work.

However, HEP has depended on the circulation of preprints since at least the mid-20th century, although the pre-Web history of the preprint infrastructure remains sketchy (Borgman 2007, Borrelli 2022, Delfanti 2016, Gunnarsdóttir 2005, Kling 2004, Till 2001, Wykle 2014). This paper examines how the preprint infrastructure originated from informal forms of communication in science, using the case of high-energy physics (HEP) and looking in particular at developments at the library of the European Organization for Nuclear Research (CERN) in the early 1960s. While libraries are often understood as archives of scholarly communication and organizations that provide communities with access to the literature, the case of the library at CERN shows how librarians actively facilitated communication in the field and constructed an infrastructure in the shadow of the formal publishing system largely by their own accord.

[1] This practice is also referred to as “green open access” (s. Taubert 2021).
[2] https://doapr.coar-repositories.org/repositories/ (accessed 21 May 2025).

Informal communication has been shown to play an important role in HEP, serving practical purposes, such as informing peers about progress and innovations in research (Gaston 1973, Libbey & Zaltman 1967, Traweek 1988). The use of arXiv.org in physics has furthermore been closely linked to the pre-existence of a "preprint culture," which emerged in the 1960s, consolidated in the 1970s, and in which researchers posted copies of their papers to colleagues (Bohlin 2004: 379, s. also Ginsparg 1994: 157). The reliance of HEP on online preprint repositories as their communication infrastructure thus reveals that "these instruments [i.e., preprint servers] are connected to specific ways of circulating scientific writing and structuring the scientific community." (Delfanti 2016: 630)

In this article, the development of preprint communication in HEP will be investigated as a process of "infrastructuring" (Barlösius 2016, 2019, Bowker & Star 1999, Star & Ruhleder 1996, van Laak 2023) the specificities of the indicated "preprint culture." The conception of infrastructure employed here emphasizes how socio-cultural aspects, such as values, expectations, or forms of sociality, are inscribed into infrastructures; and how infrastructures in turn produce forms of social structuration or solidify existing social structures that shape a social field. In other words, I try to understand how the postwar field of HEP in its social and cultural constitution influenced the organizational and technical arrangements of the preprint communication infrastructure; and how these, in turn, subsequently structured the field socially through rules and classifications.

The text conducts an archaeology that asks which practices, technologies, and organizational policies helped structure the use of preprints in HEP and solidified into the research community's early communication infrastructure. The archaeology in my case includes the description of media objects (i.e., preprints) in specific historical contexts to ask how they worked and what cultural practices and sort of sociality they facilitated (Roth et al. 2024). The aim is to understand the historical practices, technologies, institutions, and people that structured preprint communication in HEP. The investigation proceeds along three systematic-historical explorations, which all reveal features that were inscribed into the preprint infrastructure: 1) the practice of preprint communication before its formalization, particularly in postwar theoretical high-energy physics; 2) the understanding of library work and documentation at mid-century, which provided the context for constructing the preprint infrastructure at CERN; and 3) the specific case of formalizing the preprint infrastructure at the organization's library in the early 1960s.

This paper builds on my ongoing research into the history of the preprint infrastructure in HEP. It draws on archival sources as well as on historical literature pertaining to the history of physics and of scientific communication. In the next section, I will examine the origins of the “preprint culture” in HEP using the case of informal communication among a group of theorists in postwar physics to understand the socio-cultural specificities of preprint communication that became inscribed into the infrastructure. Then I will sketch the political and institutional context that defined the organization of library and documentation work at CERN, and into which the preprint culture in physics was integrated, before I describe the formalization of the preprint infrastructure at the organization. The article concludes with a discussion of what we can learn from the pre-Web history of preprint communication and its organization for a sociological understanding of scientific communication.

## 2. Face-to-Face Communication and “Preprint Culture” in Postwar HEP

The “preprint culture” in HEP, on which preprint servers in the 1990s were premised, existed since the mid-twentieth century. However, practices of “preprinting” were not wholly new to science at the time. Authors in the nineteenth century, for instance, often acquired separate print copies of their papers accepted for publication before the actual volume in which it was published came out. As Csiszar (2018: 54) notes in his history of the scientific journal, publishing papers “separately allowed authors to circulate their work more quickly,” since the publications of academies and societies appeared infrequently. Practically, preprints were thus used to overcome the problem that scientific communication was bound to specific journal release dates, ensuring that the discovery or innovation of an author become available more immediately to their peers. In the early twentieth century, the practice of distributing preprints continued in learned societies and academies, for instance, in the Faraday Society, where copies of manuscripts, which had been accepted for publication by the society’s journal, were quickly circulated to participants before a meeting (Bernal 1948a: 67).

In the twentieth-century history of physics, speedily access to new research and information has repeatedly been cited as a driving factor in the use of preprints (Coblans 1960, Gaston 1973, Goldschmidt-Clermont 2002 [1965], Kaiser 2005, Traweek 1988). The elite group of theoretical physicists that formed around Richard Feynman at Princeton’s Institute for Advanced Study in the late 1940s and early 1950s (Kaiser 2005), for instance, used preprints as an informational instrument to share their theoretical “tools” among the group without having to wait for the publication of journal papers. In this regard, Feynman’s diagram group shows some resemblance to other cases of informal communication in science, such as that of the *Drosophila* geneticists, which originated with Thomas Morgan at Caltech. This research community thrived on the active sharing of tools among members between the 1910s and 1930s – physical tools, namely fly stocks – and on information about how to use them, which helped standardize practices and “integrate geographically dispersed practitioners into a working community.” (Kohler 1994: 133) As Kohler observes: “Formal scientific publications spread the *word*, but personal exchange of working tools was probably more effective than publications, with all their elisions of craft methods, in spreading the *work* of the fly people." (ibid.) *Drosophila* researchers formalized the exchange of the "work" of geneticists through a technical newsletter, the Drosophila Information Service, which Kohler (1994: 162) calls, "in effect, an informal trade journal." Kelty (2012: 149) has additionally emphasized how the newsletter was distinguished from regular publication formats, since its purpose was to "disseminate privately" information relevant to those actively engaged in the community's work and goals. He notes that contributions to the newsletter "were not entirely restricted" but still meant to be handled "in the same manner as 'when information is obtained thru correspondence [...]'." (ibid.)

Kohler's distinction between the "word" and the "work" of a research community, as abbreviations for designating formal and informal communication, respectively, can also be used to understand preprints in postwar theoretical physics. Informal oral communication has been shown to be "fundamental to the operation of the particle physics community." (Traweek 1988: 122) As Traweek states, the main scientific work is done through discussions, while she likens "written materials, articles and preprints," to "record-keeping device[s]," whose main role it is to stake claims to priority and assign credit for discoveries or innovations (ibid., s. also Gaston 1973: 131, Bernal (1967 [1939]: 303). She explains that through talk "the boundaries of this dispersed but close-knit community" are maintained and that oral communication articulates and affirms "the shared moral code about the proper way to conduct scientific inquiry." (ibid.) However, the case of Feynman diagrams in postwar physics reveals that the distinction between oral and written communication was initially not as clear cut. Like the Drosophila Information Service in fly genetics, preprints belonged to an exchange system for spreading the "work" of theoretical physics, rather than only the word. Preprints were not simply items that added to the scientific record of theoretical HEP.

In his book on the dispersion of Feynman diagrams in postwar physics, Kaiser (2005) reveals how preprints were part of a pedagogical culture in physics in the late 1940s and early 1950s. This culture focused on postdoctoral training that "was supposed to enable young physicists to acquire those tacit skills, not codified within textbooks or quizzed on final examinations, that were essential for research." (Kaiser 2005: 65) Additionally, postdoctoral appointments in theoretical physics were usually short term and at places different from their doctoral education: "As they traveled, they picked up new skills and helped to transfer them from one place to another." (ibid.) In this context, the spread of the diagrams was simultaneously premised on personal contacts, direct instruction in using the diagrams, as well as on the distribution of preprints over far distances to compensate for the lack of face-to-face interaction and to keep up with what the group was doing. Kaiser (2005: 7ff.) argues that, rather than seeing theoretical work as removed from skill and activity, the case of Feynman diagrams shows how tacit dimensions and learned proficiency were vital to mastering theoretical tools in physics. "Feynman diagrams are *practices*, and as such they must be *practiced*. Reading about the diagrams was one thing; practicing how to use them as research tools quite another." (ibid.: 110)

Under these circumstances, physicists learned how to employ Feynman diagrams similar to how apprenticeship and practical training are needed to understand how to employ research instruments in the laboratory. Authors have shown how the work in theoretical disciplines such as mathematics and mathematical physics is highly practical and dynamic, can span several different media, and that papers and preprints are here used as a basis to manipulate and reformulate mathematical information into working instruments (Barany & MacKenzie 2014, Warwick 2003). Rather than seeing theory as transcendent and abstracted from local cultures, this means regarding the social, technical, and material conditions of theoretical work as akin to how authors have explored these dimensions regarding laboratory tools (s. Roth et al. 2024). Accordingly, many of the early diagram adopters "interacted directly with Feynman" (Kaiser 2005: 56) to either see him perform work with the theoretical tools or receive direct help in their own diagrammatic calculations.

As the postdocs who studied with Feynman and acted as his immediate "diagrammatic ambassadors" were soon scattered throughout the US and the world, however, it became vital for diagram users to keep in contact over long distances. It was important to keep abreast of the newest developments in the field, especially the newest advances in how to apply the diagrams and to learn for what purposes to use them. To this end, physicists resorted to personal correspondence through the mail, as well as other means of informal distribution, to circulate preprints and compensate for the dispersion of the community and the lack of face-to-face interaction with other practitioners. Like the genetics newsletter at the start of the century, and echoing practices of the nineteenth century, preprints were papers intended to be distributed privately to "particular friends" (Csiszar 2018: 56). In this sense, they were not meant to interfere with the regular business and social functions of academic publishing, but rather to inform researchers of the newest developments in research.

Kaiser (2005: 80) notes, though, that in the late 1940s and early 1950s, "more and more physicists began to demand preprints in order to keep up with the fast-moving research currents." Theoretical physicists at this stage maintained personal mailing lists to know exactly whom to send their papers to and with whom to inquire about the latest developments. Kaiser illustrates nicely how a leading diagram ambassador, Freeman Dyson, began soliciting from "the Institute [for Advanced Study] staff to prepare one hundred copies of the manuscript" of one of his diagram papers, "to have them ready to distribute at the New York American Physical Society meeting [in January 1949]," and how "[Robert] Oppenheimer likewise sent off preprints of Dyson's paper to a number of colleagues." (ibid.) Since preprints provided researchers with working tools, it remained customary for to maintain such private communication networks in physics well into the 1960s (Libbey & Zaltman, 1967).

## 3. The Organizational Environment and Library and Documentation Work at CERN

To understand how these private communication networks in physics could be transformed into a public information infrastructure, it is helpful to consider some of the guiding principles of CERN and its library in the early years of its existence. Briefly put, per its Convention, signed by the member states of Belgium, Denmark, France, West Germany, Greece, Italy, Netherlands, Norway, Sweden, Switzerland, UK, and Yugoslavia on 1 July 1953, CERN is an organization that is committed to international cooperation, basic research and education, and to what today might be called a policy of "openness." The founding members saw it as a duty that CERN actively promote "contacts between, and the interchange of, scientists, the dissemination of information, and the provision of advanced training for research workers." (CERN Convention 1973 [1953], Article II) This included having a welcoming stance towards scientists from anywhere in the world who wanted to conduct research on elementary particles at the Geneva laboratory. It also implied that, in principle, "membership of CERN was open to *any* state," although existing member states could veto the application of new candidates if they so pleased (Krige 1987: 221). As opposed to the original 12, CERN today has 25 member states from around the world.

Next to these political values, CERN was driven by specific academic values as well. From the experience of the war, which had still ravaged Europe less than a decade before its establishment, the organization devoted itself clearly to fostering research for civilian purposes. As the Convention states, work at the laboratory should only be "of a pure scientific and fundamental character," having "no concern with work for military requirements." (CERN Convention 1973 [1953], Article II) Its founding members thus wanted that the organization be imbued with a "scientific spirit," to prevent that intellectual life at the laboratory be dominated by the considerable attention to technical matters, which was needed for designing and building particle accelerators and detectors. In part, this was ensured by having an "eminent physicist" as the organizations Director-General as well as by establishing "a small theoretical studies group" to reside on site in Geneva (Krige 1987: 228). Krige notes further that the organization saw an example for its research characteristics in the model of the modern university. CERN was meant to encourage a pure science-attitude, attracting and supporting people who pursued scientific research "for its own sake" (ibid: 229).

The Scientific Information Service (SIS), which included the library and documentation and began operations immediately after the founding of CERN, was thus based on the organization's international outlook, commitment to basic science, and openness. It was therefore imperative that the knowledge produced at CERN "be openly published and freely discussed." (Krige 1987: 228) A 1959 "Bibliographical note on CERN publications" therefore states that, next to the commitment to fundamental and civilian research, the second most important principle of the organization is that "results of its work must be made public and generally available, thus excluding any secret or 'classified' publication." The note, which most likely was written by Herbert Coblans, first head librarian and director of the SIS at CERN from 1954 to 1964, specifies what the "policy of copen communication and wide dissemination" at the organization generally meant:

> "All publications are available, either through the exchange system or free of charge to institutions and workers in the field wherever they may be. Depositories of all publications have been established in the thirteen Member States as well as in the major accelerator centres in different parts of the world."[3]

Coblans was himself a person who was devoted to international cooperation. Before he began his work at CERN, he was head of the library mission at UNESCO with appointments promoting library ideals in South America (Carter 1978).

How such knowledge was to be made freely and widely available was, in turn, influenced by a renewed understanding of librarianship that had been forming in the first half of the twentieth century. During that period, the concept of the specialized library and the idea of documentation, that is, of indexing and cataloging specialized information for search and retrieval, were in transition. A major reason for this shift in the guiding principles of librarianship and documentation was the incredible growth of scientific research since the beginning of the twentieth century, not least accelerated by the role of science during the war (Roth 2025). While the centuries-old issue of so-called "information overload" could be met with the development of new bibliographic techniques in earlier centuries (Blair 2010), it required institutional answers in the postwar era.

Since the 1940s, librarians, documentalists, information scientists, and others were beginning to reorient the institutional values of the library. In his book *Librarianship and Documentation*, Coblans (1974: 22) describes it as an "important 'psychological' shift, namely that user needs and search habits have begun to directly influence information policy." The 1940s and 1950s saw some of the first surveys to study the habits of the specialized user community of scientists (Bernal 1948c, s. also Libbey & Zaltman 1967). In contrast, until the turn of the twentieth century what Coblans (1974: 22) calls the "keeper outlook" dominated library work and oriented it mainly towards "institutional control." Rather than constituting organizations that aimed their principles and processes for documenting the literature towards those populations, which employed it (e.g., researchers), libraries were the preserve of scholars and learned gentlemen. "Throughout the centuries it was accepted as axiomatic that scholars knew their

---

[3] n.a. "Bibliographical note on CERN publications," 17 November 1959, CERN Archive Box "SIS Library Committee pre 1964," Folder CERN-ARCH-SIS-057.

literature, and as long as libraries were well organised bibliographically, identification and access posed no serious problems." (ibid: 23) In a paper published during his tenure at CERN, he thus states that the references, which it was one of the mains functions of the library to provide, "must be intelligible and meaningful to others in many parts of the world besides the author of the paper or his immediate cronies." (Coblans 1960: 30)

The idea of the "user outlook" of the specialized library belonged to a larger set of convictions, which felt that the results of scientific research could be "applied for the common good" and that "scientific communication" forms an "essential part" in the process (Coblans 1966: 113, s. also Bernal 1967 [1939]). This view was most prominently represented by the influential Irish socialist scientist and historian J. D. Bernal, who also contributed significantly to librarianship and documentation in the 1940s and 1950s. Bernal (1948a: 54) understood that the main objectives of "scientific publication and distribution" were "the satisfaction of the user" qua scientist and "the advance of science." In his opening plenary at the Royal Society Information Conference, held from 21 June to 2 July 1948 in London, for example, Bernal summarized the current situation in scientific communication as follows:

> "What is wrong with the present system is that the growing abundance of primary scientific publication and the confusion in which it is set out acts as a continuous break, as an element of friction to the progress of science. We are not so much maintaining that scientific information is lost, though it may be, but that the scientific worker wastes time and effort in finding what information there is, and as a result we may be getting a far more limited and slower progress of research than we would under a better arranged system of publication." (ibid.)

Like Bernal, Coblans (1960: 29) states that the vast growth of science since World War II notwithstanding, "It is not for us as documentalists to question these developments; rather, we must attempt to understand the consequences." With the conviction of science as a social benefit, combined with an institutional orientation towards scientists as users, came a unique view of the active role of the library in scientific communication, which would find one form of realization in the preprint infrastructure set up at CERN. Coblans was convinced that the training and experience of "documentalists" provided them with a suitable overview of "the series of processes involved in scholarly publication," to the extent that "the techniques of recording, bibliography, photography and printing have to be integrated." (ibid.)

In line with this understanding, Bernal had proposed a radical scheme to reorganize the logistics of scientific communication at the Royal Society Information Conference. His idea anticipated some of the later orientations to scientists as users and incorporated the convention of using preprints. The proposal drew considerable backlash from conference attendants as well as outside observers, so that Bernal had to withdraw the original paper submitted to the event (East 1998). Rather than continuing to rely on societies and their journals to organize scientific communication, he suggested setting up a limited number of "National Distributing Authorities (N.D.A.)," which would coordinate the submission, distribution, and announcement of scientific papers centrally, while leaving selection and assessment of contributions largely to the panels and review boards of the appropriate academic societies (Bernal 1948b: 254). The scheme built on similar ideas he had devised in his influential book *The Social Function of Science* (Bernal 1967 [1939]: 292ff.), published in 1939, just a few months before the outbreak of World War II. Historians have likened Bernal's proposal to an "informational socialism," which is "based on the principles of centralized planning and equal access" (Aronova 2021: 134). Indeed, Bernal's ideas for the centralized organization of scientific communication influenced the All-Union Institute for Scientific and Technical Information (VINITI), which was set up by the Soviet Union's Academy of Science in 1952 (Hammarfelt & Dahlin 2023).

His scheme appropriated the idea that papers accepted for publication in a journal could be sent out for production and simultaneously be produced for quick separate circulation, thus seeing preprints as part of an information system to better keep colleagues in a field up to date. According to his plan, after having gone through the appropriate editorial stations, papers "would be at once printed by the N.D.A." and then "be published [sic] in the first place as a preprint," which would be "distributed immediately" to individuals who subscribed to papers from a certain field or requested papers announced previously, as well as be included on "a weekly list of *notices* of titles of forthcoming papers." (Bernal 1948b: 254) In this regard, Bernal merely aimed at systematizing and generalizing the already existing practice private preprint sharing and to extend it to benefit a larger population of users. Some of the procedures he devised as well as the central motif of using preprints to inform a specialized community about new developments in a field would reoccur with the formalization of preprint communication at CERN.

## 4. Formalizing and Generalizing Preprint Exchange at CERN

Against this general backdrop of librarianship and documentation, the CERN system for registering and distributing preprints turned the originally private communication networks, as they existed, for instance, in theoretical physics, into a public good for all of HEP, comprising not only theorists, but also experimental physicists, engineers, detector and accelerator constructors, and others. Much of the credit for the pioneering work of systematically registering, categorizing, and publicly announcing preprints needs to be given to Luisella Goldschmidt-Clermont, who was a librarian at CERN from after its founding until the late 1960s, when she went on to pursue a PhD in social sciences at the Free University of Brussels.[4]

Beginning in the late 1950s, the library at CERN was collecting and noting the circulation of preprints, however, not yet in an elaborate systematic scheme. In a SIS note from 1 July 1957, Goldschmidt-Clermont informed scientific staff that

> "A number of scientists in various countries have been asked to send their preprints to the CERN Library. The reason for assuming this unusual and additional task is the hope to increase the number of preprints received at CERN and to insure [sic] that they are available to all members of our scientific staff."

She asked that the CERN scientists send "4 copies" of preprints of which they were the authors to the library as well as to inform the library staff about any preprints received personally "as soon as they arrived."[5] In fact, an internal report of the SIS for the year 1957 noted that in an effort of "procuring pre-prints" some "600 scientists in 28 countries were approached," resulting in "an average 40 pre-prints" received each month since July 1957.[6] By the year 1960, an annual internal report of the SIS read that the "weekly display of pre-prints in the Library" was being "much appreciated by readers: in fact, the extent to which such single copies are requested on loan has posed problems of ensuring rapid circulation."[7]

At the end of 1960, Goldschmidt-Clermont started developing a project to formalize preprint exchange at CERN, which was based on the idea of systematically documenting and categorizing preprint submissions and making the list available in a regular newsletter that would announce the latest preprints in circulation to the local community as well as to HEP research sites around the globe.[8] When funds became available to potentially support the project in 1961, she seized the opportunity and proposed the idea to both the library administration and the CERN directorate. The rationale was as follows:

> "The project under consideration aims at improving communications among high-energy physicists [...]. A list giving author, title and abstract of papers in high energy physics and related fields received at CERN would be prepared and reproduced by offset; this list would be made widely available to scientists and laboratories doing research in high-energy physics."[9]

---

[4] https://cds.cern.ch/record/1527102?ln=en (accessed 23 June 2025).
[5] Scientific Information Service note to CERN Scientific Staff, 1st July 1957, CERN Archives Box "SIS Library Committee pre 1964," Folder CERN-ARCH-SIS-057.
[6] Scientific Information Service, "Report for the year 1957," CERN Archives Box "CERN Directors-General," Folder CERN-ARCH-DG-FILES-137.
[7] Herbert Coblans, "Report of the Scientific Information Service for the Year 1960," February 27th, 1961, CERN Archives Box CERN-ARCH-DIR-ADM-01-DIV-DD-07 to 09, Folder CERN-ARCH-DG-FILES-137.
[8] S. "CERN Library Accession List: Preprints and Reports 1969," CERN Archives Box "Directors of Administration," Folder CERN-ARCH-DIR-ADM-01-DIV-DD-07.

In contrast to the proposal made by Bernal at the 1948 Scientific Information Conference, the CERN library would thus not act as a central distribution office for preprints per se, but rather as a sort of contact exchange for interested readers and authors. Based on this list, "all those interested would know of the existence of papers being circulated on a certain subject" within one or two weeks, "and could request from the author a copy of specific papers in which they have an urgent interest." The system thus did not comprise the logistics of duplicating and distributing preprints – only a few copies were made available for display and loan at the CERN library – but only of informing the community publicly about where to request papers of interest individually. Moreover, the list was meant to include abstracts of each preprint submission, which for some members of the community could prove sufficient. Traweek (1988: 86) notes how written material like preprints was scanned by physicists "primarily to find out who to talk to." Theorists, though, required whole papers to reconstruct the documented research processes (Gaston 1973: 139).

The project idea was based on Goldschmidt-Clermont's (2002 [1965]) own informal empirical observations of "communication patterns" in HEP, which she laid out in a paper that was not published until decades later.[10] On the one hand, she saw that a general desire existed for rapid information about what was going on in the field at other sites, but that in the current situation that desire was not met satisfactorily. On the other, she argued that the moral commitments of CERN as a publicly funded and international organization demanded that this desire be realized through a publicly and openly accessible system. "Because the field is developing rapidly and because it requires large capital investments, speed is an important factor in communication. Delays occurring in the process of transmitting information may cause a waste of experimental effort." At present, "letter-journals have become the most popular journals in high-energy physics," since they often had publication delays of only "five to six weeks." (ibid.)

Letter-journals were an offspring of the "Letters to the Editors" section, which was introduced in *Physical Review* in 1932, the leading physics journal at the time (Lalli 2016).[11] The section soon took on a life of its own, expanding greatly until the mid-20th century, and later, in 1958, spawning its own publication medium *Physical Review Letters*. Already before World War II, nuclear and quantum physics were fast moving research areas and journals had tried to adapt to the quick pace of new discoveries with the introduction of the new genre of letters. The "Letters to the Editors" were meant to announce findings, which would later be backed up by full journal articles. Although the innovation was also framed as providing better access to the literature for readers, to keep colleagues up to date with the latest work, historians show how this publication format primarily constituted "a reward to the author rather than [...] a method for the readers to increase their knowledge." (ibid.: 164) Thus, while letters were meant as short announcements to be updated by full papers, Goldschmidt-Clermont (2002 [1965]) observed in 1965 that "two out of three letters in elementary particle physics have not been followed by a full paper more than two years after publication." Additionally, editors often found that letter-journals were "being misused as a channel for half-baked results or even for results that later prove[d] wrong."

---

9 Luisella Goldschmidt-Clermont memorandum to Lew Kowarski, January 5th, 1961; Luisella Goldschmidt-Clermont memorandum to Herbert Coblans, January 11th, 1963, CERN Archives Box CERN-ARCH-DIR-ADM-01-DIV-DD-07 to 09, Folder CERN-ARCH-DG-FILES-137.

10 S. also Luisella Goldschmidt Clermont letter to Viktor Weisskopf, January 4th, 1961, CERN Archives Box CERN-ARCH-DIR-ADM-01-DIV-DD-07 to 09, Folder CERN-ARCH-DG-FILES-137.

11 The idea came from the British journal *Nature*, where "the 'Letters to the editor' column" had been established earlier in the century "as a major venue for priority claims" (Baldwin 2014: 270).

Although letter-journals had been frequently employed for making fast priority claims, among HEP researchers, the format nevertheless enjoyed great popularity at the time because of the speed of publication and the concise nature of the information it contained. For Goldschmidt-Clermont this meant that, rather than understanding it as a formal publication, "the readers use[d] the letter-journals as a current-awareness tool, that is as a *communication* tool." (ibid.) From the way that researchers in HEP were utilizing the publication format she concluded that they "crave[d] a current awareness communication tool." (ibid.) In a preliminary draft of her "Proposal for the Development of Communication among High-Energy Physicists," she noted that in the prevailing situation, "the 'communication' aspect is being empirically met by distributing preprints."[12] She was referring to the sort of communication habits that characterized the group of physicists who worked with Feynman diagrams.

However, as the example of the diagram users shows, preprint communication was not an open system at the time. Although those who belonged to the close circle of users had access to the newest developments, anyone outside this circle remained virtually oblivious to what was going on until he or she could read about it, months later, in a journal publication. According to Goldschmidt-Clermont, the circulation of preprints tended to "create a privileged class: the set of scientists whose names appear on [private] mailing lists." (ibid.) Personal mailing lists were still the primary means of distribution for preprints in the mid-1960s, with more than two thirds of papers shared via such lists (Libbey & Zaltman 1967: 36). Similar situations prevailed in experimental HEP. According to an entry in a physics handbook cited by Coblans (1960: 30), a "large part of the detailed literature of proton synchrotrons is found in the privately circulated reports. These reports are available in only a few laboratories, and are not referenced because they cannot be consulted by many readers." As essentially "'unpublished' literature," he notes, these reports are "not part of the main body of the scientific record." (ibid.)

[12] Luisella Goldschmidt-Clermont memorandum to Viktor Weisskopf, January 24th, 1961, CERN Archives Box CERN-ARCH-DIR-ADM-01-DIV-DD-07 to 09, Folder CERN-ARCH-DG-FILES-137.

Accordingly, the main intention of Goldschmidt-Clermont and her preprint newsletter project was to "extend to all [researchers in HEP] the privilege which is, at present, enjoyed by the few who receive preprints."[13]

Moreover, private communication networks conflicted with the official policies of CERN, according to Goldschmidt-Clermont, which, as already indicated, prohibited any forms of secrecy as a lesson from the role of science, especially physics, during World War II. She invoked the organization's Convention in a January 1961 memorandum, stating that "CERN is bound to contribute to 'international cooperation in nuclear research, … This cooperation may include … the promotion of contacts between … scientists, the dissemination of information, …' (CERN Convention, Article II, para 3 c)." Rather than continuing the privileged system of preprints as private current awareness communication tools, she conceived of her project as inclusive of all researchers in the global HEP community. Echoing the organizations policy of openness and international outreach, Goldschmidt-Clermont explained that she was of "the opinion about the [preprint] project of high-energy physicists belonging to as many countries as possible; some of these physicists are staff members [at CERN]; some were only on a short visit to CERN; a few of the latter are responsible physics leaders in their countries."[14]

Thus, the main service realized through the preprint infrastructure set up by Goldschmidt-Clermont at CERN was the reliable and regular provision of "current awareness communication tools" to the entire HEP community. Having requested that authors send copies of their "preprints, lecture notes, reports, etc."[15] to the library, she ensured that the entries would be catalogued and put on public display for the local research community as well as included in a newsletter. Goldschmidt-Clermont declared: "No discretion would be made between what is currently called preprints, reports, internal reports, private communication, etc[.]." That the library treated preprints as "communication tools" also meant that "No attempt would be made to select the papers according to the scientific value of the work presented therein." Instead, criteria for inclusion were tied to formal rules, which defined what had to be included in the submission ("the full paper should be supplied with the abstract; the abstract should be in English; the full paper could be in any language.") as well as the categorization of a paper.[16] All submissions as preprints would be cataloged uniformly with a unique preprint number, regardless of what precise format they were.

[13] Luisella Goldschmidt-Clermont memorandum to Lew Kowarski, January 5th, 1961, CERN Archives Box CERN-ARCH-DIR-ADM-01-DIV-DD-07 to 09, Folder CERN-ARCH-DG-FILES-137.
[14] Luisella Goldschmidt Clermont letter to Viktor Weisskopf, January 4th, 1961, CERN Archives Box CERN-ARCH-DIR-ADM-01-DIV-DD-07 to 09, Folder CERN-ARCH-DG-FILES-137.
[15] Scientific Information Service note to CERN Scientific Staff, 1st July 1957, CERN Archives Box "SIS Library Committee pre 1964," Folder CERN-ARCH-SIS-057.
[16] Luisella Goldschmidt-Clermont memorandum to Lew Kowarski, January 5th, 1961, CERN Archives Box CERN-ARCH-DIR-ADM-01-DIV-DD-07 to 09, Folder CERN-ARCH-DG-FILES-137.

Goldschmidt-Clermont noted that the "general reaction is one of enthusiasm, with no doubt about the usefulness of such a project." The main reason why virtually all physicists gladly complied with the new procedure was because it rectified some crucial issues in the current communication infrastructure: "The physicists I have spoken to feel that our project may help clarify – rather than deteriorate – the sensitive matters of priorities, adequate quotations, etc."[17] In the issue of priorities, Goldschmidt-Clermont reported about the divergence between the date of submission and of publication of a journal article, an issue that had also plagued publishing in the nineteenth and early twentieth century. While the latter was naturally taken as the date to establish priority, the factual date of priority was that of submission.[18] The submission of preprints to the CERN library and registration through the list of entries in a public newsletter "would help to disclose reals issuance dates." Effectively, the CERN registry thus became a place to clearly stake claims to priority within the field, though these claims were not necessarily universally accepted outside of the tight-knit HEP community (Llewellyn Smith 2023). Additionally, Goldschmidt-Clermont noted, it was customary in the current system "to refer to other people's work still available only in preprint form."[19] Consequently, a system, which would not only register issuance dates, but also identified submissions in standardized form would essentially ensure that the papers could be unambiguously cited.

However, without intention, the preprint system at CERN sowed the seeds for developing social functions akin to those found in formal publishing systems. The library set clear criteria for the categorization of submissions: Preprints dealing with "theoretical particle physics" and "high energy experimental physics" were always included. "Experimental techniques" related to HEP, "accelerators," "mathematics," "cosmic rays," and "computer programmes" constituted subjects that would be "considered for inclusion in the newsletter," depending "on the case." "Low energy physics," "chemistry," and "plasma physics" were, in turn, never included.[20] The enforcement of these strict boundaries already foreshadowed some of the rigid categorizing that the sorting software and moderators of arXiv.org perform today, which also do not see their job in evaluating the scientific quality of paper but nevertheless have a considerable impact on the structure of HEP research (Reyes-Galindo 2016). Moreover, though these categories appear as neutral classifications of the field and its boundaries, they were slanted towards the interests of the local community. As Goldschmidt-Clermont explained, the CERN community at the time served as a proxy for determining the relevance of submitted papers: "The guiding principle is that the [preprints] list should only include 'hot' material, that is material for which it pays to use this exceptional procedure. In order to determine if it 'pays,' one has to consider the number of staff members [i.e., researchers at CERN] interested in the subject [...]."[21]

---

[17] Luisella Goldschmidt Clermont letter to Viktor Weisskopf, January 4th, 1961, CERN Archives Box CERN-ARCH-DIR-ADM-01-DIV-DD-07 to 09, Folder CERN-ARCH-DG-FILES-137.

[18] Csiszar (2018, 161) discusses how in the mid-19th century conflicts persisted over whether the date at which a paper was presented to a society (and accepted for publication) or whether the publication date of a journal issue constituted the proper coverage of priorities: "Until mid-century, it was not regularly asserted that priority claims depended directly on putting those claims into print."

[19] Luisella Goldschmidt Clermont letter to Viktor Weisskopf, January 4th, 1961, CERN Archives Box CERN-ARCH-DIR-ADM-01-DIV-DD-07 to 09, Folder CERN-ARCH-DG-FILES-137.

[20] Luisella Goldschmidt-Clermont memorandum to Herbert Coblans, January 11th, 1963, CERN Archives Box CERN-ARCH-DIR-ADM-01-DIV-DD-07 to 09, Folder CERN-ARCH-DG-FILES-137; Scientific Information Service, "CERN Library Accession List: Preprints and Reports 1969," CERN Archives Box "Directors of Administration," Folder CERN-ARCH-DIR-ADM-01-DIV-DD-07.

It is thus already possible to glimpse at how the infrastructure moved from providing current information to structuring the community socially. Through its public preprints announcement, CERN became the center of gravity in the organization of preprint communication as its newsletter constituted an (incomplete) representation of research in the field. Inclusion in the list was therefore a badge of membership as well as an opportunity to expose one's work to the global HEP community in efforts to boost one's reputation. However, the success of one's exposure and the chances to receive credit hinged on conforming with the subject classifications and thus with the research interests and "hotness" of subjects define at CERN. This social incentive seems to have predetermined a reward system, which later became institutionalized in the arXiv online repository, as Delfanti (2016: 642) explains: "The archive is a detector that HEP physicists need to hit with their beams of papers, at the right time and with the right objects, in order to signal their existence and claim credit for what they have done." At CERN, it was not "the archive" but the library's preprints list that had to be hit with paper beams in a similar manner.

## 5. Conclusion

The idea of scientific communication is still dominated by views, which see it as a domain controlled largely by scientists, particularly in their changing roles as authors, editors, reviewers, and readers of the academic literature. Scientific communication is thus presented in mostly formal terms in sociological studies: the system of (journal) publication organizes credit and reward and fulfils essential gatekeeping functions (e.g., Bourdieu 1975, Latour & Woolgar 1982, Zuckerman & Merton 1971). As a result, the common understanding of scientific communication "focuses on social mechanisms like peer review, the recognition of merit or the attribution of reputation," while the role of the organizations and infrastructures that facilitate scientific communication remains overlooked (Taubert 2021: 176). In an age where communication in science is defined by online publishing platforms and large multinational conglomerates, the sociological understanding of scientific communication needs an update.

[21] Luisella Goldschmidt-Clermont memorandum to Herbert Coblans, January 11th, 1963, CERN Archives Box CERN-ARCH-DIR-ADM-01-DIV-DD-07 to 09, Folder CERN-ARCH-DG-FILES-137.

The archaeology of the pre-Web preprint infrastructure in HEP can help refine the understanding of scientific communication. It emphasizes the role organizations such as specialized libraries can play in actively maintaining communication in scientific fields. The library at CERN constructed an infrastructure that incorporated socio-cultural specificities of communication in postwar physics into a system borne by organizational values oriented to international collaboration, openness, and user needs. Through the work of Goldschmidt-Clermont, and under the auspices of Coblans, the library turned existing informal and private communication networks into a centrally organized system for the public distribution of preprints and therefore into a good benefitting the whole HEP community. The system helped standardize communication in the field by introducing a preprint registry and newsletter at the CERN library, which became widely available. It thus allowed members within HEP to treat preprints somewhat like publications, making them citable and acting as records of priority claims within the community. While the release of information about the newest preprints was now bound to regular intervals determined by the library, it nevertheless alleviated the trouble caused by long publication time-lags and irregular release dates.

However, despite its self-image as primarily a public "information service" to the HEP community and the sincere commitment to neutrality regarding the scientific content of submitted papers, the case of the CERN library also shows how the infrastructure did not act neutrally. Instead, the decisions and routines that shaped the information system and oriented it to the research interests of the laboratory in Geneva biased the way the accession of preprints was handled. Moreover, although the preprint infrastructure was erected, as it were, in the shadow of conventional academic publishing, as the system scaled and become more established, it gradually also acquired social functions of scientific communication usually associated with formal publishing. These findings lend themselves to revising our sociological understanding of scientific communication.

The archaeology of the pre-Web preprint infrastructure overcomes assumptions about a supposed neutrality of the media technologies and infrastructures that communication in science is based on. Rather, my archaeology shows how organizations such as libraries, tasked with facilitating a specialized community's access to information, act like sociotechnical systems whose values and institutional logics determine the shape of a public communication infrastructure. What makes the idea of preprint communication so appealing in the eyes of many today, is that it allows circumventing traditional institutions of academic publishing. Accordingly, commentators have described preprint communication as "scientists taking the dissemination process and the ownership of papers into their own hands" (Gunnarsdóttir 2005: 556). However, the CERN case can help better understand how even without formal institutions like peer review the infrastructures and organizations, which facilitate communication in science, exhibit socially structuring capacities shaped by the routines, values, and politics that underly their technologies (Bowker & Star 1999). While the practice of self-archiving on online platforms is seen to boost the autonomy of scientific authors, my study supports research that shows how power over the shape of scientific communication is exerted by its infrastructures, whether maintained by libraries or by online platforms (Reyes-Galindo 2016). Generally spoken, a sociological understanding of scientific communication should take heed of the sociotechnical systems that act as platforms and infrastructures of communication and publication in science, including libraries and preprint servers, but also the smaller and very large publishing houses, and ask how they exert powers that bias the way scientific discourses are structured and limited.

**Acknowledgements:**

The author wishes to express his gratitude to David Dallman, Salomé Rohr, and Jens Vigen of the CERN library and archive for their generous help and genuine interest in the project. Further thanks go to Arianna Borrelli, Stefan Böschen, and Matthew Eisler for useful comments on an earlier version of the text. The research for this paper was supported by the Käte Hamburger Kolleg: Cultures of Research, RWTH Aachen University, funded by the German Ministry of Research, Technology, and Space (BMFTR).